\documentclass[prl,twocolumn,superscriptaddress,showpacs,floatfix]{revtex4-1}
\usepackage{epsfig}
\usepackage{color}
\usepackage{graphics}
\usepackage{amsmath}
\newcommand{ \gA} {\it {\bf A}}
\newcommand{ \gB}  {\it {\bf B}}
\newcommand{ \gal}  {\bf \alpha}
\newcommand{ \galp}  {\bf \alpha'}

\begin{document}

\title{Epigenetic Dynamics of Cell Reprogramming}

\author{S. S. Ashwin}\email{ss.ashwin@gmail.com}
\affiliation{Department of Computational Science and Engineering, Nagoya University,
Nagoya 464-8603, Japan}

\author{Masaki Sasai}\email{ sasai@cse.nagoya-u.ac.jp}
\affiliation{Department of Computational Science and Engineering, Nagoya University,
Nagoya 464-8603, Japan}

\date{\today}

\begin{abstract}
Reprogramming is a process of transforming differentiated cells into pluripotent stem cells by inducing specific modifying factors in the cells. Reprogramming is a non-equilibrium process involving a  collaboration at levels separated by orders of magnitude in time scale, namely transcription factor binding/unbinding, protein synthesis/degradation, and epigenetic histone modification. We propose a model of reprogramming by integrating these temporally separated  processes and show that stable states on the epigenetic landscape should be viewed as a superposition of basin minima generated in the fixed histone states. Slow histone modification is responsible for the narrow valleys connecting the pluripotent and differentiated states on the epigenetic landscape, and the pathways which largely overlap with 
these valleys explain the observed heterogeneity of latencies in reprogramming. We show that histone dynamics also creates an intermediary state observed in experiments. A change in the mechanism of histone modification alters the pathway to bypass the barrier, thereby accelerating the reprogramming and reducing the heterogeneity of latencies.
\end{abstract}

\maketitle
Mammalian {\it differentiated} cells having specialized functions in the adult body are generated from  fertilized egg cell. This differentiation process was thought to have defined a physiological arrow of time and was considered irreversible. 
A paradigm shift occurred when Takahashi and Yamanaka \cite{Yamanaka1}  demonstrated that differentiated mouse cells can be reprogrammed to {\it induced pluripotent stem cells} (iPSC) by inducing certain factors (known as Yamanaka factors (YF):  Oct4, Sox2, Klf4, and c-Myc)  in the cell. iPSC can differentiate into a variety of specialized cells, paving  way for a revolution in medical sciences~\cite{Jaenisch}. However, a major challenge remains; reprogramming  is inefficient, so that only a small portion of cells infected with YF transform to iPSC, and a microscopic understanding of the mechanism is the need of the hour. Insightful clues come from the quantitative analyses by Hanna et al. \cite{Hanna}, which indicated that reprogramming exhibits distributed latencies, where the latency is defined as time required for a YF infected differentiated cell ({\it founder cell}) to generate a daughter iPSC. This indicates stochasticity \cite{Yamanaka2} and hence a role for a statistical physics analysis of reprogramming \cite{Morris,JWang,PWang,Sasai2013}.

A statistical physics model  needs to incorporate epigenetic histone dynamics, which has not been considered explicitly in previous models of reprogramming.
Unlike simple bacterial genes, gene expression in eukaryotes is orchestrated by the formation of loosely and tightly packed chromatin structures, where the former is termed {\it euchromatin} and the latter  {\it heterochromatin}. DNA is typically wrapped around a protein complex  known as a {\it histone octamer}. In order for gene expression to proceed, the DNA should be unwrapped in the euchromatin structure, so that RNA polymerase and other factors can access binding sites on the DNA \cite{Maeshima}. 
The modification of histones and their subsequent interactions with DNA determine the chromatin structure \cite{Sneppen2012, Hathaway}. Modification of histones through methylation is heritable, invoking a heritable gene activity above the DNA level, termed as {\it epigenetics}. 
\begin{figure}[t]
\includegraphics[width=2.3in]{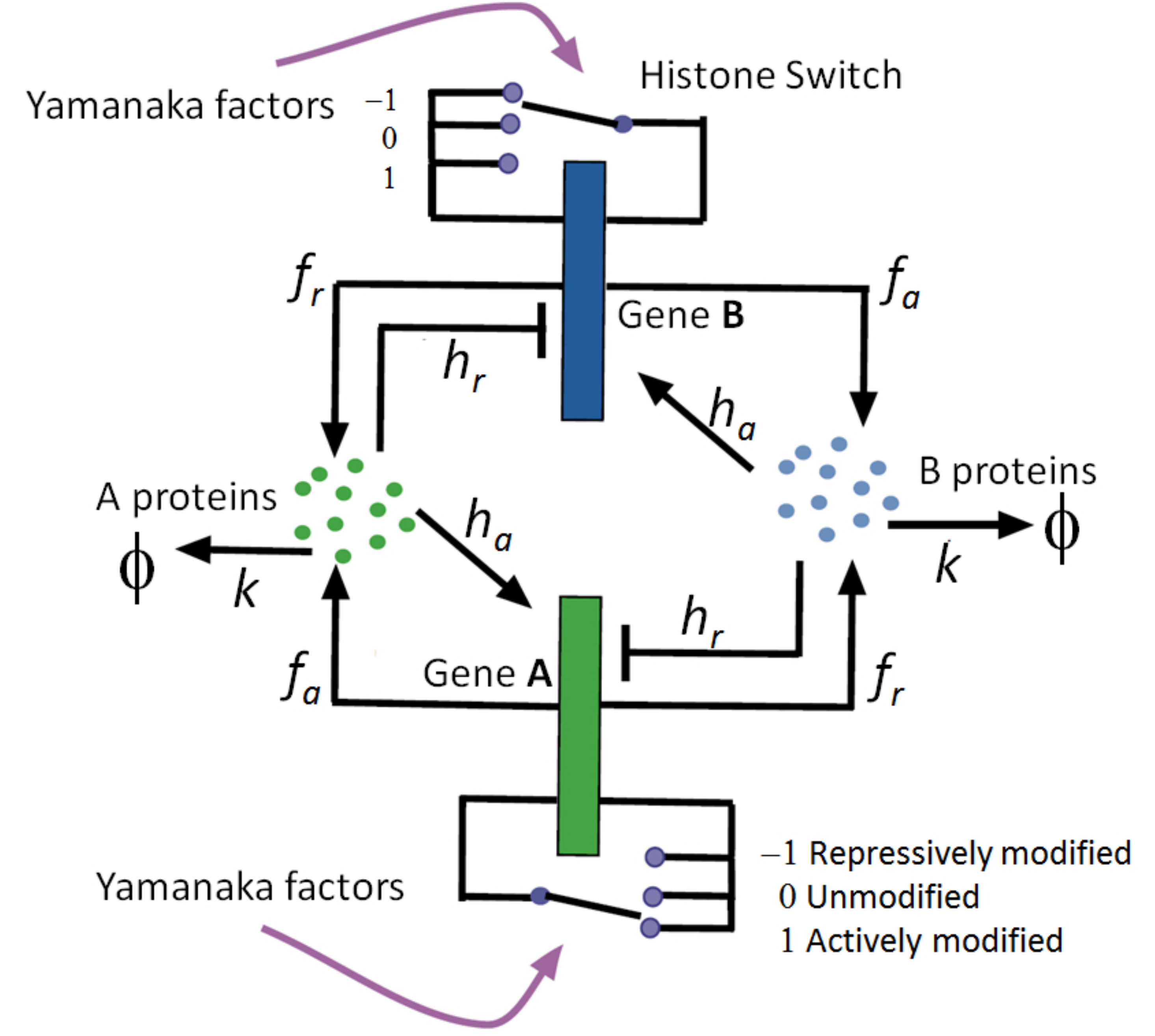}
\caption{A schematic of the repressor-activator gene regulatory network in connection with the three state histone switch. The blunt ends denote repression and the pointed ends to the gene represent activation. Binding of repressor leads histone switch to the repressive state and binding of activator leads histone switch to the active state. YFs bind as pioneer factors \cite{Soufi2012} to change the HS.}  
\label{fig:network}
\end{figure}
Epigenetic modifications play a crucial role in reprogramming, so that the iPS, differentiated, and  intermediary cells are in different epigenetic states. 
YF modify the epigenetic state of a differentiated cell in order to convert it to an iPSC \cite{Papp}. 
Thus, the explicit theoretical treatment of epigenetic histone dynamics is necessary. Here, 
we introduce a model integrating mechanisms at the histone level (slow time scales) and transcription factor binding/unbinding (fast time scales).

A particular emphasis will be placed on the quantification of landscape that characterizes stability of cells and pathways of transition between cell types, {\it i.e.}, the epigenetic landscape (EL) \cite{Waddington,Goldberg,Zhang}. We show that slow histone dynamics creates new low-barrier pathways on the EL, which are used by the reprogramming mechanism. At an ensemble level, the heterogeneity of latencies was modulated experimentally \cite{Hanna,Rais2013}, but the mechanism of this modulation is not known \cite{Zviran}. In the present work, we show that the change in epigenetic dynamics on the EL should be a key mechanism to elucidate this problem.

\begin{figure}[t]
\includegraphics[width=2.9in]{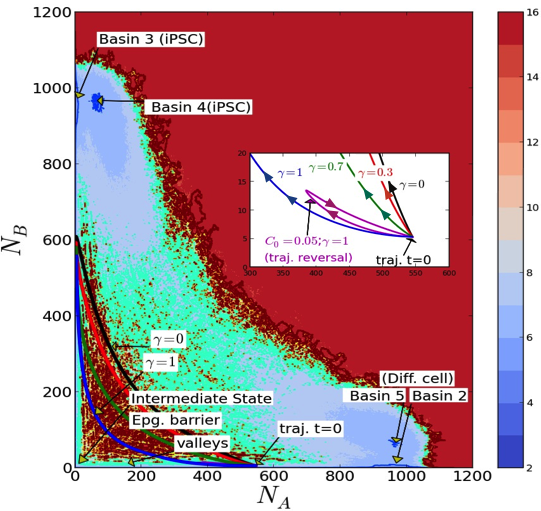}
\caption{Trajectories of $(\langle N_A(t)\rangle, \langle N_B(t)\rangle)$ drawn on the steady state landscape for a repressor-activator switch integrated with the HS modification; 
(i) differentiated and iPSC states distributed around basins at $N_A>>N_B$ and those at $N_B>>N_A$, respectively, 
(ii) appearance of narrow valleys with low barriers (epigenetic valleys) that connect the intermediate state to other states, (iii) trajectory with $\gamma=1$ largely overlaps with the epigenetic valleys, but trajectories with decreasing $\gamma$ (0.7, green), and (0.3, red) depart from the valleys, and the trajectory with $\gamma=0$ (black) bypasses the epigenetic barrier.  The trajectory with $\gamma=1$ and $C_0=0.05$ (magenta) reverts to the differentiated state. $C_0=0.1$ for other trajectories and $\tau =100k^{-1}$ for all trajectories.}  
\label{landscape}
\end{figure}

We start with a  multi-stable gene regulatory model without histone modification dynamics~\cite{Enver}, namely the repressor-activator network model introduced by Huang and coworkers \cite{Huang2005,Huang2009,Huang2010}. 
In the repressor-activator network, the  protein produced by one gene represses the other gene, but positively regulates its own expression. The state vector in this model is ${\bf N}(t)=(N_A(t),N_B(t))$, where $N_A$ and $N_B$ are copy numbers of proteins $A$ and $B$ synthesized from genes $\gA$ and $\gB$, respectively. EL is defined by $-\log[P_s(N_A,N_B)]$ in the $N_A$-$N_B$ space, where $P_s(N_A,N_B)$ is the steady state probability distribution. $\gA$ and $\gB$ work in an antagonistic way to represent the switching transition between the $N_A >>N_B$ and $N_A<<N_B$ states. This A-B network motif is ubiquitous in regulating differentiation as 
{\it Oct4-Cdx2} and {\it Nanog-Gata6}, for example \cite{Ralston05,Orkin08,Loh08}. Though this network model was used originally to describe the selection between two lineages \cite{Huang2005,Huang2009,Huang2010}, we highlight the inclusion of histone dynamics which plays a vital role in eukaryotic gene expression. We regard $\gA$ as a marker gene specific to a differentiated cell and $\gB$ as a pluripotency gene such as {\it Nanog}, which is specific to iPSC, so that reprogramming is the transition from the differentiated cell with $N_A >>N_B$ to the iPSC with $N_A<<N_B$. 
 
%
In eukaryotes, the chromatin structural change plays significant roles, which are described here with the coarse grained representation of the histone state (HS). We assume that the state of a chromatin region around each gene locus, which includes a few hundred histone octamers, is collectively denoted as $s=-1,0$, or $1$~\cite{Xing2014}: (i) In the $s=-1$ state, histones are repressively modified, (ii) in $s=0$, unmodified, and (iii) in $s=1$, actively modified.
In terms of structure, $s=-1$ is heterochromatin, $s=1$ is euchromatin, and $s=0$ is also heterochromatin but ready to form a euchromatin state. We should note that although the modification reaction of individual histones is quick and histones in chromatin and those in nucleoplasm are frequently exchanged, the change in HS representing the cooperative change of many histones occurs on a time scale of a week accompanied by dynamical DNA methylation/demethylation \cite{Hathaway,Zohar}. Change in the protein copy number through translation/transcription and degradation, on the other hand, occurs at time scale of several hours \cite{Thomson} showing the large gap of characteristic time scales between two processes. 
Thus, the transition of chromatin, $s=-1\, {\rightarrow}\, 0$ or $s=0\, {\rightarrow}\, 1$  (and vice-versa), is a slow switching mechanism  \cite{Hathaway}. See Fig.\ref{fig:network} for the illustration of the model. The gene is active only when $s=1$. Protein $A(B)$ is an activator of gene $\gA (\gB )$ and a repressor of gene $\gB (\gA )$. When a repressor binds and deactivates the gene, the repressor-binding state is set to 0 (or OFF), on unbinding it is turned  1 (or ON). Similarly when an activators binds, the activator-binding state is set to 1 and on unbinding set to 0 (or OFF). The entire  network state is then defined by the number of proteins $N_A,N_B$ and the $24$ states of gene $\gA$ and $\gB$ {\it i.e.} $s=-1,0,1$, repressor ON/OFF, activator ON/OFF denoted by $\mid N;\text{ } s=-1,0,1; \text{ repressor state } j=0,1; \text{ activator state } m=0,1\rangle$. 
We can now think of the state vector ${\bf N}(t)$ as a trace over a subspace of the gene states {\it i.e.,}  ${\bf N}=\sum_{s_{A,B}, j_{A,B}, m_{A,B}} (\, \mid N_A\, s_A\, j_A\, m_A\rangle,\,\, \mid N_B\, s_B\, j_B\, m_B\rangle\, )$.  Since the laboratory observable state is  ${\bf N}$, the HS remain as  hidden variables as far as the conventional EL is concerned. The trajectory on the EL is an average over the HS.

For simplicity, we assume $\gB$ and $\gA$ are symmetric having the same parameters.
The rates and time scales in the model are in units of $k$ and $k^{-1}$, respectively; $k\approx 0.1\, {\rm h}^{-1}$ \cite{Thomson}  being the protein degradation rate, and length of a cell cycle is about $2k^{-1}$ \cite{Hanna} though we do not include cell cycle explicitly. 
At a given gene state $\mid N_{\alpha} s j m\rangle$ with $\alpha=$ A or B, the production rate of protein is $g_{sjm}$. The parameters are biologically motivated \cite{parameters}; when $s=1$, the activator is bound (ON) and the repressor is unbound (ON), the protein production of the gene is maximal and is denoted by $g$. Repressor binding always reduces the protein production rate, hence the other rates are fractions of $g$. When the $s=-1$ or $0$, the protein production rate is set to 0. We assume proteins bind to DNA in a dimer form for simplicity \cite{Zhang}, so that the  rates of binding are $h_{a}N_{\gal}(N_{\gal}-1)$ and $h_{r}N_{\gal}(N_{\gal}-1)$. We set $f_{a}/h_{a}=f_{r}/h_{r}=50000$ to make the ratio $f_{a,r}/(h_{a,r} N_{\gal}^2)<1$ for a typical protein level in an eukaryotic nucleus  $N_{\gal}\approx O(10^3)$ with the unbinding rates $f_{a}=f_{r}=10k$. The rates of HS change are defined in terms of $\{q_l,q_l^{'},r_l,r_l^{'}\}$ with $l=1,..,4$. The governing stochastic equations are given in \cite{set1,set2,set3}. The rate of HS switching is set to much smaller than the rate of protein number change as $q=0.05k$. This corresponds to a time scale  $q^{-1}=20k^{-1}\approx 1$ week \cite{Hathaway}. We assume positive feedback relations between protein synthesis and histone modification; the HS tends to be turned active when the activator binds, and turned repressive when the repressor binds. We, therefore, have $q_1>>q_4$ and $r_1^{'}<<r_4^{'}$ \cite{parameters2}.
\begin{figure}[t]
\includegraphics[width=3.1in]{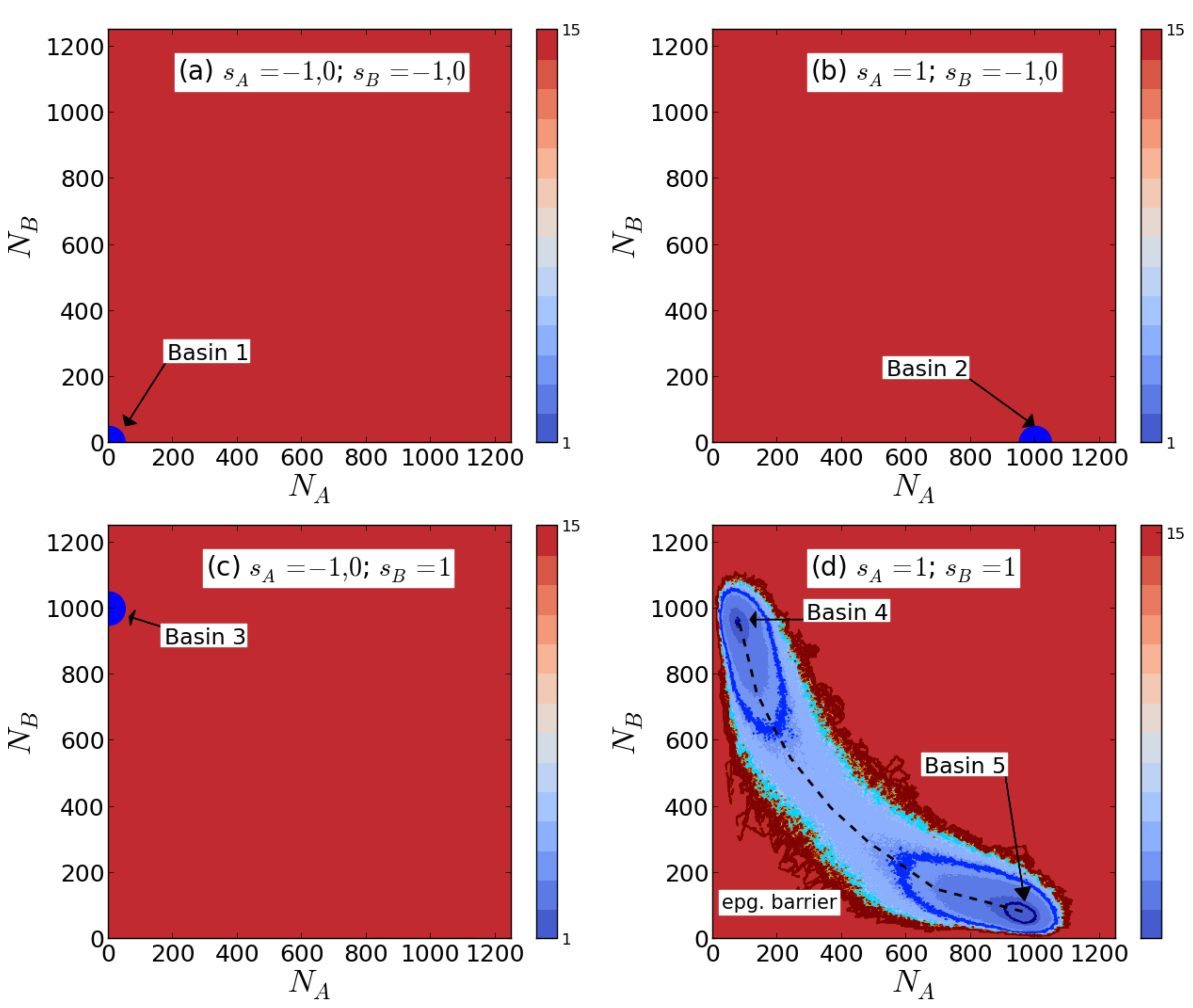}
\caption{Landscapes at the fixed HS: (a) $s_{A}=s_{B}=0$, (b) $s_{A}=1, s_{B}=0$, (c) $s_{A}=0, s_{B}=1$, and (d) $s_{A}=s_{B}=1$. A dashed line connecting basins in D is drawn as a guide for eyes.}
\label{superposition}
\end{figure}

We first calculate the EL: $-\log P_s(N_A,N_B)$ using the Gillespie algorithm~\cite{Gillespie} for the stochastic equations \cite{set1,set2,set3} (Fig.~\ref{landscape}), where the steady state probability $P_s(N_A,N_B)$ is calculated using 100 trajectories each over $10^8$ time steps long with random initial conditions.  EL shows five basin minima, each of which corresponds to a steady state solution at a fixed HS;
when HS are OFF ($s=0,-1$) for both the genes, the model has a basin at $N_A=N_B=0$ (basin 1, Fig.~\ref{superposition}a), while $\gal$ gene HS is ON the other OFF, the solution corresponds to $N_{\gal}\neq 0$ the other protein number being zero (basin 2, Fig.~\ref{superposition}b; basin 3, Fig. \ref{superposition}c). When both histones are ON ($s=1$), the model has two basins (basins 4 and 5), which are the same solutions as in Huang's model \cite{Huang2005,Huang2009,Huang2010} (Fig.~\ref{superposition}d).
Thus, in the presence of slow switching between HS, we have pathways connecting basins as 1-2, 2-5, 1-3, 3-4, and 4-5. Distribution over basins 3 and 4, distribution over basins 2 and 5  and distribution concentrated around basin 1 define the iPSC, differentiated and intermediate states, respectively. 
Epigenetic dynamics creates these low-barrier valleys between the basin minima as shown in Fig. \ref{landscape}, which are not present in Huang's model. We will show, that trajectories with large latency distribution tend to use these valleys.

Our approach is to determine the evolution trajectories of the system via the master equation. The probability distribution $\vec{P}(N,t)$ is a 24 dimensional vector with components $(P_{A,111}(N_A,t),...P_{B,-100}(N_B,t))$, with indices running first for $A$ then $B$ in the following sequence  $111,110,101,100,011,010,001,000,$ $-111,-110,-101,-100$.  The master equation then is:
\begin{eqnarray}
\frac{ d\vec{P}(N,t)}{dt}&=&{\bf G} \left(\vec{P}(N-1,t)-\vec{P}(N,t)\right) \\ \nonumber
&+& k\left((N+1)\vec{P}(N+1,t)-N\vec{P}(N,t)\right) \\ \nonumber
&+& \left( {\bf F+H+Q+R}\right)\vec{P}(N,t) +{\bf C}\vec{P}(N,t).
\label{eqn:master}
\end{eqnarray}
Protein generation matrix {\bf G} is diagonal with elements $\{g_{111},...g_{-100}\}$. The scalar $k$ is a degradation term.   {\bf F} and {\bf H}~\cite{MatFH} represent unbinding and binding of proteins from/to genes,  and {\bf Q} and {\bf R}~\cite{MatQR} are the HS transition matrices. 
The matrix {\bf C} represents the effects of YF.

When YF are induced in the cell, they tend to transform the HS.  Since the precise action of YF is not known, we interpolate between two possible mechanisms; (I) YF work as histone-mark erasers by changing the HS as $s_A=1\rightarrow 0$  and $s_B=-1\rightarrow 0$, and (II) they work also as activators on {\gB} as $s_A=1\rightarrow 0$ and $s_B=-1\rightarrow 1$.  We here consider that these two mechanisms work with the relative importance factor $0\le \gamma\le 1$. Here, $\gamma =1$ when YF solely act as histone-mark erasers, and $\gamma =0$ when they are efficient to activate the HS in {\gB}. Thus, the Yamanaka matrix is ${\bf C}=C(t)(\gamma{\bf C}_{\rm I}+(1-\gamma){\bf C}_{\rm II}))$, here ${\bf C}_{\rm I}$ and ${\bf C}_{\rm II}$ are matrices representing the above mechanisms I and II~\cite{MatC}, and $C(t)=C_0\exp(-t/\tau)$ is the effectiveness of YF with $\tau$ being the lifetime of ectopic expression.
We first relax the system to the differentiated state with $C_{0}=0$ and then let the the system relax with $C_{0}\neq0$ from time $t=0$.

\begin{figure}[t]
\includegraphics[width=2.9in]{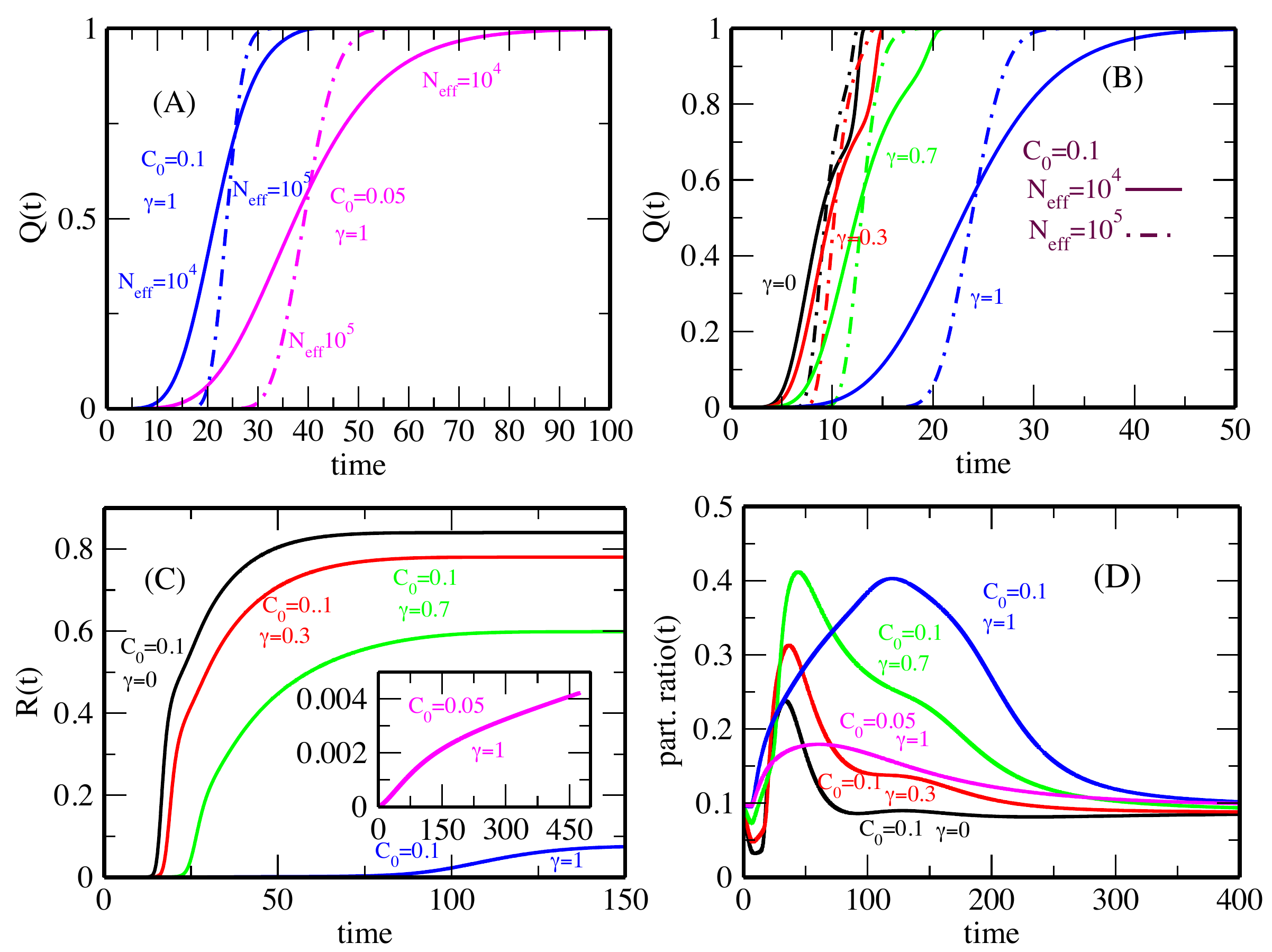}
\caption{Reaction kinetics of reprogramming with various working mechanisms and efficiencies of Yamanaka factors. (A) Probability of a founder cells to generate daughter iPSC, $Q(t)$, with $\gamma =1$ for $C_0=0.05$ and $0.1$, (B) Q(t) with $C_0=0.1$ and different $\gamma$, (C) probability of individual cells to reach the iPSC state, $R(t)$, with inset $C_0=0.05$
and the rest $C_0$=0.1. (D) Participation ratio to estimate the degree of localization of the distribution $P(N_A,N_B,t)$.   $N_{\rm eff}=10^5$ (solid line) or $10^4$ (dash and dot line) and $N_{\rm thr}=5\times 10^{-4}N_{\rm eff}$ in A, B  and $\tau=100k^{-1}$ in all panels.}
\label{cumulative}
\end{figure}

We solve the master equation under the proteomic field approximation (PFA) \cite{pfa}, which considerably reduces the dimensionality of the master equation \cite{Sasai2003, Walczak2005}.
We start by analyzing the average protein numbers $\langle N_{\alpha}(t)\rangle=\sum_{N_{\gal},i,j,m} N_{\gal}P(N_{\gal},i,j,m)$ for $\alpha =$A and B. As shown with $(\langle N_{A}(t)\rangle, \langle N_{B}(t)\rangle)$ in Fig.~\ref{landscape} for the case of $\gamma=1$ (blue trajectory), starting from the $t=0$ point near the differentiated basin, the system first proceeds along a valley of $N_B\approx 0$ and surpasses the epigenetic barrier near the intermediate state. We have $\langle N_{A}\rangle\approx \langle N_{B}\rangle\approx 0$ around the intermediate state, which is consistent with the observed late activation of the pluripotency genes after the lineage specific genes being repressed \cite{Brambrink2008,Buganhim2012}. 
On crossing the epigenetic barrier, the trajectory finds a pathway along the other valley of $N_A\approx 0$ to reach the iPSC state. We should note that these flat valleys emerge due to slow epigenetic dynamics, and are absent in the previous models of gene switches which neglected dynamics of histone modification. By decreasing $\gamma$, the trajectory  departs from epigenetic valleys (green and red), and with $\gamma =0$,  the trajectory bypasses the epigenetic barrier (black) suggesting the rapid reprogramming is realized along this pathway.


At an ensemble level, the calculated epigenetic dynamics of reprogramming can be compared with experiments  \cite{Hanna,Rais2013}. In the experiments, $N_{\rm col}$ founder cells infected with YF were placed in $N_{\rm col}$ wells on a plate at $t=0$ to multiply and form genetically identical clones. Population of these cells in each well exponentially increased from 1 to $10^6$ to reach a steady state in $t> 10$ days \cite{Hanna}. The signature of an iPSC is the expression of Nanog. The probability $Q(t)$, that a daughter iPSC is generated from a founder cell,  was estimated from the observed number, $N_{\rm nanog^{+}}(t)$, of colonies that contained  Nanog expressing cells at time $t$. One then has,  $Q(t)=N_{\rm nanog^{+}}(t)/N_{\rm col}$. A first principle estimation of $Q(t)$ is obtained from the model at an ensemble level.

Let $N_{\rm eff}$ be the effective population size of a colony and $N_{\rm thr}\approx 1$ define the minimum  threshold number of cells to label a well as iPSC detected. $R(t)$ is the cumulative fraction of iPSC in this ensemble of cells with $R(t)=1-P(t)$, where the survival probability $P(t)=\sum_{N_A,N_B}P(N_A,N_B;t)$ is obtained by solving the PFA equation with an absorbing boundary condition in the iPSC state \cite{note1}. Assuming the cells in the effective population of differentiated cells can be regarded as independent, we can write the fraction of colonies generating iPSC: $Q(t)=\sum_{n> N_{\rm thr}} \frac{N_{\rm eff}!}{n!(N_{\rm eff}-n)!} R(t)^nP(t)^{N_{\rm eff}-n}$.

Fig.~\ref{cumulative}a shows $Q(t)$ for various $C_0$ and $N_{\rm eff}$ \cite{note2} with the mechanism $\gamma =1$. For both cases of $C_0=0.1$ and 0.05, $Q(t)\approx 0$ in the initial phase and starts to rise at $t_0$ and reaches $Q(t)\approx 1$ at $t_1$ showing that colonies had heterogeneously distributed latencies. For $C_0=0.05$, we have $t_0\approx 28k^{-1}$  and $t_1\approx 95k^{-1}$, which agrees with the experimentally observed data, $t_0 \approx 30k^{-1}$ and $t_1\approx 100$-$200k^{-1}$. As shown in Fig.~\ref{cumulative}b, slope of $Q(t)$ becomes larger as $\gamma$ decreases. With $\gamma =0$,  $Q(t)$ increases much more rapidly with $t_0\approx 8k^{-1}$ and $t_1\approx 10k^{-1}$ for $N_{\rm eff}=10^5$, which is similar to the observed data with $t_0\approx 8k^{-1}$ and $t_1\approx 12k^{-1}$ obtained for cells in which Mbd3, a factor which binds to the methylated region of DNA, is silenced \cite{Rais2013}. Thus, when YF work as histone-mark erasers, reprogramming has heterogeneous latency distribution, but when YF work also as activators of pluripotency genes, reprogramming is accelerated with lesser degree of heterogeneity or is more ``deterministic'' \cite{Hanna,Rais2013} in latencies. Increased heterogeneity in latency  for the case $C_0=0.05$ can be accounted for by the reverting trajectory (Fig 2). The trajectory is unable to cross the epigenetic barrier and reverts, due to the low concentration of YF. During this process, the tails of the distribution are absorbed in the iPSC sink at  a rate which depends on the distance between the peak of the distribution and the sink, creating a large latency distribution.


Difference in heterogeneity of latencies between two cases is also found by plotting $R(t)$ as in Fig.~\ref{cumulative}c; increase of $R(t)$ is much sharper in the $\gamma =0$ case than in  $\gamma =1$. Difference between two cases becomes further evident when we plot the participation ratio, $\sum_{N_A,N_B}P(N_A,N_B;t)^2/P(t)^2$, which is large when the distribution $P(N_A,N_B;t)$ is localized in the $N_A$-$N_B$ space and small when it is delocalized. Fig.~\ref{cumulative}d shows that the distribution is more localized in the $\gamma =1$ case during the reprogramming, showing that population is accumulated around the intermediate state. Localization pattern is found to be more complex in the $\gamma =0$ case, which should be experimentally detectable by the single-cell level tracking during reprogramming. 


We have introduced a simple model for reprogramming by integrating the histone modification mechanism with the gene expression mechanism, providing a consistent view on kinetics of reprogramming and the stability of cell states.
We have elucidated how pathways are determined on the EL aided by histone modification dynamics. Models of this kind will provide details of the trajectory and barriers helping experimentalists with microscopic information which is otherwise difficult to obtain in order to build efficient schemes for reprogramming.
It is important to apply concepts and methods developed here to more realistic networks \cite{JWang,PWang,Sasai2013,Zhang} involving  larger number of pluripotency and lineage specific genes.



\begin{thebibliography}{999}
\bibitem{parameters}
$g_{111}=g$, $g_{110}=0.8g$, and $g_{101}=0.1g$. For all other $sjm$, we set $g_{sjm}=0$. In order to make copy numbers order of  $\sim100's$, which is typical of eukaryotes, $g$ is set to 1000.
\bibitem{note1} 
The absorption condition was imposed as $P(N_A, N_B,t)=0$ for $N_B>910$, and $P(N_A, N_B,t)$
was multiplied by a factor $1-\exp(-(910-N_B)/100)$  for $760<N_B\le 910$.
\bibitem{note2}
Because the HSs are inherited from mother to daughter cells, there should be correlation among multiple cells in a colony. The effective number of cells, $N_{\rm eff}$, therefore, should be smaller than the actual number of cells in a colony. We here used $N_{\rm eff}=10^4$ and $10^5$.
\end{thebibliography}
\end{document}